\documentclass[aip,prb,twocolumn,reprint,showpacs,spuerscriptaddress]{revtex4}
\usepackage{graphicx}
\usepackage{color}
\usepackage{amsmath}
\usepackage{epsfig}
\usepackage{lineno}

\newcommand{\vgap}{\vspace{0.1cm}}

\newcommand{\konl}{k_{\rm{on}}^{\rm L}}
\newcommand{\koffl}{k_{\rm{off}}^{\rm L}}
\newcommand{\kecl}{k_{\rm{ec}}^{\rm L}}
\newcommand{\kons}{k_{\rm{on}}^{\rm S}}

\newcommand{\kecs}{k_{\rm{ec}}^{\rm S}}
\newcommand{\Dl}{D_{\rm L}}
\newcommand{\Ds}{D_{\rm S}}
\newcommand{\gl}{g_{\rm L}}
\newcommand{\gs}{g_{\rm S}}
\newcommand{\gp}{g_{\rm P}}
\newcommand{\kp}{k_{\rm P}}
\newcommand{\CA}{C_{\rm A}}
\newcommand{\SA}{S_{\rm A}}
\newcommand{\cA}{c_{\rm A}}
\newcommand{\sA}{s_{\rm A}}
\newcommand{\Cl}{C_{\rm L}}
\newcommand{\Cs}{C_{\rm S}}
\newcommand{\cl}{c_{\rm L}}
\newcommand{\cs}{c_{\rm S}}

\begin{document}

\title{A mathematical model of intercellular signaling during epithelial wound healing}

\author{Filippo Posta}
\affiliation{Dept. of Biomathematics, UCLA, Los Angeles, CA 90095-1766}
\email[]{fposta@ucla.edu}
\author{Tom Chou}
\affiliation{Dept. of Biomathematics, UCLA, Los Angeles, CA 90095-1766} 
\affiliation{Depts. of Biomathematics and Mathematics, UCLA, CA 90095}
\email[]{tomchou@ucla.edu}

\date{\today}

\begin{abstract}
Recent experiments in epithelial wound healing have demonstrated the
necessity of Mitogen-activated protein kinase (MAPK) for coordinated
cell movement after damage. This MAPK activity is characterized by two
wave-like phenomena. One MAPK ``wave'' that originates immediately
after injury, propagates deep into the cell layer, and then rebounds
back to the wound interface.  After this initial MAPK activity has
largely disappeared, a second MAPK front propagates slowly from the
wound interface and continues into the tissue, maintaining a sustained
level of MAPK activity throughout the cell layer.  It has been
suggested that the first wave is initiated by reactive oxygen species
(ROS) generated at the time of injury.  In this paper, we develop a
minimal mechanistic diffusion-convection model that reproduces the
observed behavior.  The main ingredients of our model are a
competition between ligand (e.g., Epithelial Growth Factor) and ROS
for the activation of Epithelial Growth Factor Receptor (EGFR) and a
second MAPK wave that is sustained by stresses induced by the slow
cell movement that closes the wound.  We explore the mathematical
properties of the model in connection with the bistability of the MAPK
cascade and look for traveling wave solutions consistent with the
experimentally observed MAPK activity patterns.
\end{abstract}

%\keywords{MAP kinase, reactive oxygen species, traveling waves}
\pacs{87.10.Ed, 87.17.Aa.}

\maketitle

% \linenumbers

%% main text
\section*{Introduction}
Coordinated cell movement is an essential feature of many biological
processes, such as wound healing, embryonic morphogenesis, and tumor
growth \citep{Martin04}. In wound healing, cell migration and cell
contraction are the two main mechanisms responsible for wound
closure. Cell contraction is the dominant mechanism in the closing of
small wounds through the so called ``purse-string" process
\citep{Kiehart99}.  For larger wounds, cell contraction is not
sufficient, and surrounding cells must migrate to close larger wounds.
While the two mechanisms are not mutually exclusive, there are cases
where cell migration is the only healing process \citep{Sherratt90},
such as when a strip of cells from an epithelial layer is removed
\citep{Poujade07}. Despite the existence of experimental assays
targeting cell migration during wound healing, there are still many
open questions.  For instance, before injury, the cells are resting,
but after wounding they become motile. What mechanical and biochemical
phenomena regulate motility?  Is it the availability of free space
that leads cells to move toward wound closure?  What determines the
speed of cell migration?  To be able to answer these questions, we
need to understand both the mechanical and biochemical aspects of cell
migration, and how they might regulate each other.  While the physical
mechanisms of cell movement have been well-studied \citep{DiMilla91,
Sherratt90,Murray03,Maini04}, the complex regulation of the wound
healing process by biochemical signals and feedback pathways remains
poorly understood.  Recent experimental investigations of wounds in
epithelial tissue have highlighted novel properties of the
intercellular signaling necessary for healing
\citep{Matsubayashi04,Nikolic06}.

Matsubayashi {\it et al.} (2004) analyzed wounded epithelial
monolayers of Madin-Darby canine kidney (MDCK) cells and showed
coordinated movement not only by the cells at the wound edge, but by
several rows of adjacent cells.  In their experiments, cell
proliferation did not play a significant role.  Their study also
showed that the cellular response of the epithelial monolayer is
qualitatively characterized by two wave-like activation patterns of
ERK 1/2, a Mitogen Activated Protein Kinase (MAPK).  These ``waves"
are characterized by a time-dependent front of higher ERK 1/2
concentration that initiates at the wound edge and spreads into the
cell layer.  The first rebounding wave-front propagates into the
tissue and back to the wound quickly, the second wave-front is slow
and sustains MAPK activity in the tissue until wound closure. This
second, final wave appears to be related to cell migration through a
positive feedback loop, since inhibition of the second MAPK wave halts
coordinated cell movement.  Moreover, during this second ``wave'',
MAPK is inactive around the healing edges of the injured layer, but
still active in the migrating cells around the open wound
\citep{Matsubayashi04}.

Nikoli\'{c} {\it et al.} (2006) extended the experimental analysis in
\citep{Matsubayashi04} by probing the epithelial wound healing assay
with a novel wound generating technique. They used
polydimethylsiloxane (PDMS) slabs to create two different wounding
protocols.  A peel-off injury is created by growing part of the
epithelial monolayer over a PDMS slab. Once the slab is peeled-off,
taking with it the cells grown over the PDMS, it creates a wound that
breaks some cells at the wound edge, but leaves the space surrounding
the wound free of cellular debris. The other protocol consists of the
removal of a PDMS slab that forms the boundary of the cellular
monolayer, its removal leaves intact the cells in direct contact with
the slab, but now provides free space at one edge of the monolayer.
This ``unconstraining'' protocol avoids cell tearing upon PDMS
removal.  Experimenting with these two techniques, and with the
standard scratch wound assay, the authors confirmed the presence of
two separate waves of MAPK in case of injury.  The unconstraining
experiments expressed only the slow, final wave.  In this case, cell
movement was limited and random, suggesting that availability of free
space is not enough to generate an organized migration of the
epithelial layer and that mechanical injury is necessary.
Accordingly, scratch and peel-off experiments resulted in both waves
of MAPK activation and in collective monolayer migration.  Using
immunofluorescence, Nikoli\'{c} {\it{et al.}} (2006) were also able to
identify reactive oxygen species (ROS) as one of the key components of
the intercellular signaling responsible for the observed pattern of
MAPK activation. They discovered that ROS are generated immediately
after injury and remain present around the wound edge for at least the
duration of the first rebounding wave.  Inhibition of ROS by
N-acetyl-L-cysteine resulted in the absence of both MAPK waves and
cell migration. The results from the two studies
\citep{Matsubayashi04, Nikolic06} are summarized in Table
\ref{tab:ressummary}.

In this paper, our aim is to build a mathematical model that
reproduces an that can be used to analyze the properties of MAPK
activation during wound healing experiments.  Because of its relevance
and ubiquity, the MAPK pathway remains the subject of many
computational and mathematical modeling studies \citep{Orton05}.  The
MAPK cascade is a signal transduction pathway that relays an external
stimulus to the cell, it is characterized by the sequential activation
of three protein kinases \citep{Ferrell96}.  Huang and Ferrell
\citep{Huang96} proposed a system of 18 differential equations
representing the 10 reactions that compose the three-kinase MAPK
cascade.  A mathematical and computational analysis of the system
showed that the cascade has the effect of amplifying an input signal
(e.g., receptor phosphorylation) in such a way that its overall
behavior can be compared to that of a cooperative enzyme
\citep{Keener98}. These first computational studies sparked additional
modeling efforts that, coupled with ongoing discoveries by
experimentalists, have led to many more advanced models that exhibit
characteristics (e.g., bistability, ultrasensitivity, oscillations,
etc.) of the MAPK signaling pathway \citep{Qiao07,
Hornberg05,Schoeberl02,Sasagawa05}.
 
Here, we are not interested in the intracellular dynamics of MAPK, but
rather on its role within the wound healing signaling network.  For
this reason, we will treat the MAPK cascade as a black box, using the
results in \citep{Ferrell96,Ferrell97tbs} to essentially represent the
whole cascade as a switch for signal transmission.  In this approach,
MAPK/ERK is the output of the ``signaling switch''.  ROS are one input
that can activate the switch, but are unlikely to be the only one
because of the different properties of the two activation waves.  As
suggested in \citep{Nikolic06} and in other wound healing experiments
\citep{Xu04,Block04}, other inputs are diffusible ligands and their
cell receptors. For instance, Epidermal Growth Factor (EGF) and EGF
Receptor (EGFR) play essential roles in promoting cell migration,
proliferation and wound closure \citep{Wiley03,Joslin07}.  Moreover,
positive and negative feedbacks between EGFR signaling and the MAPK
cascade have been demonstrated experimentally and verified
computationally \citep{Santos07,Kholodenko06,Kholodenko07}.  Although
Nikoli\'{c} {\it et al.} suggest EGF as a possible signal, as well as
other molecules, they did not pursue the topic in their work. However,
they did identify Reactive Oxygen Species (ROS) as direct regulator of
MAPK activity in their wound healing experiments.  Since ROS have been
shown to induce EGFR activation in the absence of EGF
\citep{Reynolds03}, this finding is in agreement with other studies
that showed the regulatory role of ROS in wound healing
\citep{Roy06,Sen08} and MAPK signaling \citep{Torres03,McCubrey06}.

Because of the documented connection between diffusible signals, wound
healing and MAPK activation, we propose a mechanistic model based on
ligand-mediated intercellular signaling that reproduces the observed
MAPK activation pattern, and that is consistent with the qualitative
experimental features listed in Table \ref{tab:ressummary}.

\begin{table}[tb]
\begin{center}
\begin{tabular}{c | c}
\hline
Experiment & Results  \\
\hline
Scratch wound 
% \citep{Matsubayashi04}  
& \parbox[c]{2in}{\vgap Two MAPK waves, cell migration\vgap}  \\
 Closed wound 
 %\citep{Matsubayashi04}  
 &  
 \parbox[c]{2in}{\vgap No MAPK activity, no cell movement\vgap} \\
 MAPK inhibition 
 %\citep{Matsubayashi04}  
 &  
 \parbox[c]{2in}{\vgap No MAPK waves, no cell migration\vgap} \\
 Slow wave inhibition 
%\citep{Matsubayashi04}  
 & 
 \parbox[c]{2in}{\vgap No cell migration\vgap} \\
%\hline 
Peel-off wound 
%\citep{Nikolic06}  
& 
\parbox[c]{2in}{\vgap Two MAPK waves, cell migration\vgap} \\
 Unconstraining 
 %\citep{Nikolic06} 
 &  
 \parbox[c]{2in}{\vgap Slow, second wave only, no cell migration\vgap} \\
 ROS inhibition 
 %\citep{Nikolic06} 
 &  
 \parbox[c]{2in}{\vgap No MAPK waves, no cell migration\vgap}\\
\hline
\end{tabular}
\end{center}
\caption{Summary of the experimental results from \citep{Matsubayashi04, Nikolic06}.}
\label{tab:ressummary}
\end{table}

\section*{Mathematical Model}

\begin{figure}[tb]
\begin{center}
\includegraphics[width=8cm]{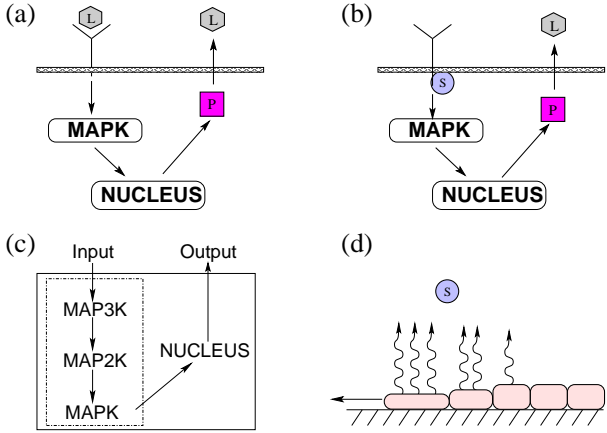}
\end{center}
\caption{Schematic of biophysical events during wound healing. (a) diffusible ligands (L)
phosphorylate membrane receptors by binding to them, activating the MAPK
cascade that leads to the production of intracellular protease (P). Protease 
induces ligand release. 
(b) ROS (S) can also interact with membrane  
receptors by inducing phosphorylation of  their cytoplasmic tail, 
which also activates the MAPK cascade and promotes release of
ligands into the extracellular matrix. 
Diffusible ligand and ROS represent two independent triggers of the MAPK cascade.
(c) Schematic of the three-kinase cascade
that characterize MAPK signaling. 
(d) The stresses caused by cell movement toward wound
closure can lead to ROS release  \citep{Rhee00, Torres03, Ali06}.}
\label{fig:signaling}
\end{figure}

The experimental results from \citep{Matsubayashi04, Nikolic06}
provide evidence of spatio-temporal MAPK signaling for the regulation
of cell migration during wound healing, without determining the exact
biochemical events that govern it.  The fact that free space by itself
is not enough to produce coordinated cell migration suggests that ROS
is not the only diffusible signal needed to generate the two waves of
MAPK activation. To explain the observed spatio-temporal MAPK pattern,
at least two diffusible signals are needed.  Both ROS and EGF
molecules are able to diffuse in the extracellular space (ROS can also
move across the cell membrane) and both can phosphorylate the EGF
membrane receptor (EGFR), activating the MAPK cascade.  EGF induces
phosphorylation of the cytoplasmic tail of EGFR by binding to it.
Reynolds {\it et al.} (2003) showed that ROS can induce EGFR
phosporylation even in the absence of EGF by binding to intracellular
phosphatases. Also well documented is the positive feedback between
EGF and the MAPK cascade, and its ability to produce long range
signaling through autocrine relays \citep{Pribyl03}.  Finally, ROS can
be generated by mechanical stresses like the ones generated by
migration of the epithelial monolayer \citep{Torres03}, thus providing
a feedback loop between MAPK activation, cell motility, and further
ROS production. These four signaling mechanisms are summarized in
Fig. \ref{fig:signaling} and constitute the foundation of our
mathematical model.

The wound healing system is a three-dimensional one, and the migration
of individual cells toward wound closure is not exactly normal to the
wound edge as shown by cell tracking experiments \citep{Nikolic06}.
Here, we simplify the analysis by considering the cell layer in cross
section as a semi-infinite straight line, with the wound initially
positioned at the origin.  The resulting two-dimensional system
consists of a semi-infinite cell layer that is immersed in medium of
infinite height (the medium in \citep{Nikolic06} is 3mm which is much
larger than any other length scale in the problem).  We model four
species: ROS, one diffusible ligand ({\it e.g.}, EGF), one ligand
receptor ({\it e.g.}, EGFR) and playing the role of the output of the
MAPK cascade ``black box", a protease that is the intracellular
precursor of the ligand ({\it e.g.}, the piece completing the feedback
loop between EGF and the MAPK cascade in Figs.
\ref{fig:signaling}-(a) and \ref{fig:signaling}-(b)). We denote the
local concentrations of these species by $L$, $S$, $R$, and $P$
respectively. As depicted in Fig. \ref{fig:signaling}, the signal can
be transmitted to the cell in two different ways: through a
ligand-receptor complex ($\Cl \equiv R \cdot L$) and a ROS-receptor
complex\footnote{ROS are known to activate EGFR by associating with
its cytoplasmic tail, and inactivating its phosphatase
activity. Kinetically, modeling this process is equivalent to
irreversible ROS-EGFR ($\Cs \equiv R \cdot S$) complex formation.}
($\Cs \equiv R \cdot S$). We will assume that the number of available
cell membrane receptors is in excess, implying that $R$ is
approximately constant and that ROS-ligand-receptor complexes are
negligible.

\begin{figure}
\begin{center}
\includegraphics[width=8cm]{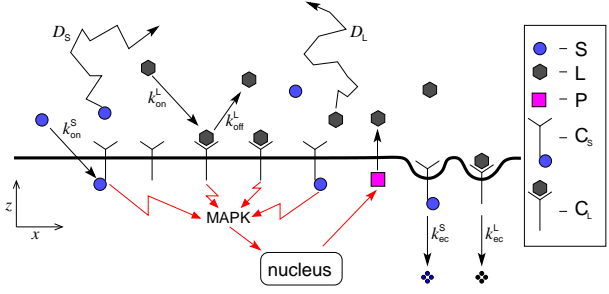}
\end{center}
\caption{Schematic of simplified signaling pathway during  
wound healing: ligands ($L$) and ROS ($S$) are free
to diffuse and bind to cell surface receptors. This binding forms complexes
($\Cs$ and $\Cl$) that activate MAPK signaling. Complexes are lost due to endocytosis or,
in the case of ligand-receptor complexes, through ligand unbinding and 
endocytosis. Intracellular proteases ($P$) release extracellular ligands ($L$).}
\label{fig:Framework}
\end{figure}

A schematic representation of the system is given in Fig.
\ref{fig:Framework}. The governing equations and boundary conditions
of our model are

\begin{widetext}
\begin{eqnarray}
\frac{\partial L(X,Z,T)}{\partial T} &=&  \Dl \left( 
\frac{\partial^2 L(X,Z,T)}{\partial X^2} + \frac{\partial^2 L(X,Z,T)}
{\partial Z^2} \right), 
L(X,Z=\infty,T) = 0, \label{eqn:L} \\
\Dl \frac{\partial L(X,0,T)}{\partial Z} 
 &=& \konl R L (X,0,T)-\koffl \Cl(X,T)- 
\gl P(X,T), \label{eqn:Lbc} \\
%&&
\frac{\partial \Cl(X,T)}{\partial T} &=&  \konl R L(X,0,T) - 
\left( \koffl + \kecl \right) \Cl(X,T), \label{eqn:CL} \\
%&&
\frac{\partial S(X,Z,T)}{\partial T}  &=&  \Ds \left( 
\frac{\partial^2 S(X,Z,T)}{\partial X^2} + \frac{\partial^2 S(X,Z,T)}
{\partial Z^2} \right), 
%\hspace{0.25cm} 
S(X,Z=\infty,T) = 0, \label{eqn:S} \\
%&&
\Ds \frac{\partial S(X,0,T)}{\partial Z}
&=&  \kons R S (X,0,T) - \gs \Pi_{\rm S}(\Cl,X,T), \label{eqn:Sbc}\\
%&&
\frac{\partial \Cs (X,T)}{\partial T}  &=&  \kons R S(X,0,T) - 
\kecs \Cs(X,T), \label{eqn:CS} \\
%&&
\frac{\partial P(X,T)}{\partial T}  &=&   - \kp P(X,T)+ \gp \: \Pi_{\rm P}\left( \Cl,\Cs
\right). \label{eqn:P}
\end{eqnarray}
\end{widetext}
Equations (\ref{eqn:L}) and (\ref{eqn:S}) describe the diffusion of
ligands and ROS in the extracellular medium with homogeneous diffusion
constant $\Dl$ and $\Ds$, respectively.  Eqn. (\ref{eqn:Lbc}) accounts
for the flux of ligand across the surface of the cellular layer
including ligand-receptor complex formation with rate constant
$\konl$, ligand-receptor complex dissociation with rate constant
$\koffl$, and extracellular ligand release by intracellular protease
with rate $\gl$.  Eqn. (\ref{eqn:CL}) governs the kinetics of
ligand-receptor complexes, with new complexes forming at rate $\konl$
and dissociating with rate $\koffl$, $\kecl$ represents the rate of
receptor-mediated endocytosis of the ligand-bound receptor complexes.
Eqn. (\ref{eqn:Sbc}) describes the diffusive flux of ROS due to
formation of ROS-receptor complexes with rate $\kons$ and to ROS
production by the functional $\Pi_{\rm S}(\Cl,X,T)$.  The kinetics of
ROS-receptor complexes is represented by Eqn. (\ref{eqn:CS}) and its
terms are analogous to the ones in Eqn. (\ref{eqn:CL}), except that
there is no ROS release from the ROS-receptor complex in accordance
with \citep{Reynolds03}.  The last equation describes the cellular
response to extracellular signaling through the activity of
intracellular proteases. Within the wound healing framework, protease
activity is directly related to ERK1/2 activity measured in
\citep{Nikolic06}. In particular, the protease dynamics is
characterized by a degradation term with rate constant $\kp$ and a
source term $\Pi_{\rm P}(\Cl,\Cs)$ with maximum production rate $\gp$.
To complete the description of the mathematical model, we need to
impose reasonable functional forms for $\Pi_{\rm P}$ and $\Pi_{\rm
S}$.

The role of the functional $\Pi_{\rm P}(\Cl,\Cs)$ is to represent the 
intermediate biochemical steps that
lead to protease production. These steps include 
the MAPK cascade and any other reaction
in the feedback loop between ligand binding and ligand release ({\it e.g.}, 
the solid box in Fig. \ref{fig:signaling}-(c)). In the literature,
$\Pi_{\rm P}$ is usually represented as a sigmoidal function of cell surface complexes
 such as the Hill function \citep{Ferrell96,Ferrell97tbs,Shvartsman02}. 
If the level of receptor 
signaling is given by the total concentration of complexes ($\Cl + \Cs$), 
we propose the following functional form:
\begin{equation}
\Pi_{\rm P}(\Cl,\Cs)=\frac{\left( \Cl +\Cs \right)^n} {\CA^n + \left(
\Cl +\Cs \right)^n},
\end{equation}
where $n$ is an effective Hill coefficient, 
and $\CA$ represents an activation threshold of the signaling pathway.

Defining the ROS source $\Pi_{\rm S}$ is more problematic since the
experimental evidence suggests an interplay between cellular signaling
and cell migration, thus involving mechanical forces whose description
goes beyond the scope of this paper.  Conversely, the production of
ROS due to wound induction is embedded in the initial conditions of
the system and it is not described in $\Pi_{\rm S}$.  Generally, ROS
production increases with cell metabolism \citep{Torres03}.  In our
case, metabolic increase may be related to cells becoming motile
\citep{Ali06, Torres03} and/or to ligand signaling \citep{Rhee00}.  We
use a Hill function multiplied by a decaying exponential to represent
$\Pi_{\rm S}$:
\begin{equation}\label{eqn:JDim}
\Pi_{\rm S}(\Cl,X,T)=\left\{ \begin{array}{cl} 0 & T < T_{\rm D} \\
\frac{\left(\Cl(X,T-T_{\rm D})\right)^m}
{\SA^{m} + \left(\Cl(X,T-T_{\rm D})\right)^m} \; e^{-k_x X} & T \geq T_{\rm D} 
\end{array} \right. . 
\end{equation}
The Hill functional represents ligand-mediated ROS production stemming
from the phosphorylation of a receptor's tail during ligand
binding. The delay $T_{\rm D}$ represents the delay between ligand
binding and ROS production.  Finally, the exponential factor in Eqn.
(\ref{eqn:JDim}) describes reduction in ROS production due to the
decrease in motility from cells near the wound edge ($X=0$) to cells
farther from it.

Upon introducing the following dimensionless quantities
\begin{eqnarray}
&& t=\kp T, \hspace{0.2cm} x=X\sqrt{\frac{\kp}{ \Dl} }, \hspace{0.2cm} z=Z\sqrt{\frac{\kp}{ \Dl} }, 
\nonumber \\[13pt]
&& l=L \frac{ \kp \kecl \konl R}{ \gl \gp \left(\koffl + \kecl \right)}, 
\hspace{0.2cm} \cl= \Cl \frac{ \kp \kecl}{\gl \gp}, \hspace{0.2cm}
\cs= \Cs \frac{\kecs}{\gs}, \nonumber \\[13pt]
&& s= S \frac{\kons R}{\gs}, \hspace{0.2cm} p= P \frac{\kp}{\gp},\nonumber
\end{eqnarray}
we express the system of equations in dimensionless form:
\begin{eqnarray}
\frac{\partial l(x,z,t)}{\partial t} & = &   
\frac{\partial^2 l(x,z,t)}{\partial x^2} + \frac{\partial^2 l(x,z,t)}
{\partial z^2},
%\hspace{0.25cm} \lim_{z \rightarrow \infty} s(x,z,t)=0; 
\label{eqn:Lnd} \\[13pt]
\alpha \; l_z (x,0,t) & = & l (x,0,t)- \nonumber \\
& - &\beta \left[ l(x,0,t)-\cl(x,t)
\right] - p(x,t), \label{eqn:Lbcnd} \\[13pt]
\varepsilon \; \frac{\partial \cl (x,t)}{\partial t} & = & l(x,0,t) - 
\cl(x,t), \label{eqn:CLnd} \\[13pt]
\delta \; \frac{\partial \cs (x,t)}{\partial t} & = & s(x,0,t) - 
\cs(x,t), \label{eqn:CSnd} \\[13pt]
\frac{\partial s(x,z,t)}{\partial t} & = & \eta \left( 
\frac{\partial^2 s(x,z,t)}{\partial x^2} + \frac{\partial^2 s(x,z,t)}
{\partial z^2} \right),  
%\hspace{0.25cm} \lim_{z \rightarrow \infty} s(x,z,t)=0; 
\label{eqn:Snd} \\[13pt]
\nu \; s_z (x,0,t) & = & s(x,0,t) - \pi_{\rm s}(\cl,x,t), \label{eqn:Sbcnd} \\[13pt]
\frac{\partial p}{\partial t} & = &  -p + \pi_{\rm p}\left( \cl,\cs
\right), \label{eqn:Pnd}
\end{eqnarray}
where the dimensionless parameters are 
%\begin{eqnarray}
\begin{equation}\label{eqn:ndpar}
%&&
\begin{array}{c}\displaystyle
\alpha=\frac{\sqrt{\Dl \kp} \left( \koffl +\kecl \right)}
{\konl R \kecl}, \hspace{0.2cm} \beta=\frac{\koffl}{\kecl},
\hspace{0.2cm} \delta=\frac{\kp}{\kecs}, \\[13pt]
%&& 
\displaystyle
\varepsilon=\frac{\kp}{\koffl+\kecl}, \hspace{0.2cm} \eta=\frac{\Ds}{\Dl},
\hspace{0.2cm} \nu=\sqrt{\frac{\kp}{\Dl}} \frac{\Ds }{\kons R}. \end{array}
\end{equation} 
%\end{eqnarray}
The parameters $\alpha$ and $\nu$ characterize the relative rates of
diffusion and binding, while $\beta$ represents the strength of
complex degradation relative to ligand dissociation. The parameters
$\varepsilon$ and $\delta$ describe the speed of binding and
endocytosis relative to ligand release mediated by intracellular
species, and $\eta$ is the ratio of diffusivity between ROS and EGF
ligand.

The dimensionless protease and ROS production functions become
\begin{eqnarray}\label{eqn:sigmand}
\pi_{\rm p}(\cl,\cs)=\frac{\left( \cl +\gamma \: \cs \right)^n}
{\cA^n + \left( \cl +\gamma \: \cs \right)^n}, \hspace{3cm}
\\ 
\pi_{\rm s}(\cl,x,t)=\left\{ \begin{array}{cl} 0 & t < \tau \\
\displaystyle 
\frac{\left(\cl(x,t-\tau)\right)^m}
{\sA^{m} + \left(\cl(x,t-\tau)\right)^m}e^{-\lambda x} & t \geq \tau
\end{array} \right.,
\end{eqnarray}
where $\cA= \CA ( \kp \kecl)/(\gl \gp)$, 
$\gamma=(\gs \kp \kecl)/(\gl \gp \kecs)$, $\tau=T_{\rm D} \kp$, 
$\lambda=k_x L$, and $\sA=(\SA \kp \kecl)/(\gl \gp)$.

\subsection*{Fast Binding Approximation}

Before we proceed to the analysis of the model, 
we are going to make an assumption that significantly simplifies the model and that 
is also justifiable biophysically. 
We assume that ligand dissociation and complex degradation 
is fast compared to protease degradation, {\it e.g.}, 
$\kp \ll \kecs,\koffl,\kecl$. 
In this limit, $\varepsilon$ and $\delta$ are small and 
Eqns. (\ref{eqn:CLnd}) and (\ref{eqn:CSnd}) 
can be treated as a singular perturbation. 
On timescales of protease degradation the concentration of ligand-receptor and 
ROS-receptor complexes are approximately that of the surface concentration of free ligand 
and ROS, respectively. If we consider only the ``outer'' solutions of 
Eqns. (\ref{eqn:CLnd}) and (\ref{eqn:CSnd}) , our full model reduces to the three equations:

\begin{eqnarray}
\frac{\partial l(x,z,t)}{\partial t} & = &   
\frac{\partial^2 l(x,z,t)}{\partial x^2} + \frac{\partial^2 l(x,z,t)}
{\partial z^2}, 
%\hspace{0.25cm} \lim_{z \rightarrow \infty} l(x,z,t)=0,
\label{eqn:lm} \\
\alpha \; l_z (x,0,t) & = & l (x,0,t)
 - p(x,t), \label{eqn:lmbc} \\
\frac{\partial s(x,z,t)}{\partial t} & = & \eta \left( 
\frac{\partial^2 s(x,z,t)}{\partial x^2} + \frac{\partial^2 s(x,z,t)}
{\partial z^2} \right), 
%\hspace{0.25cm} \lim_{z \rightarrow \infty} s(x,z,t)=0,
\label{eqn:sm} \\
\nu \; s_z (x,0,t) & = & s(x,0,t) - \pi_{\rm s}(l,x,t), \label{eqn:smbc} \\
\frac{\partial p}{\partial t} & = &  -p + \pi_{\rm p}\left( l,s
\right), \label{eqn:pm}
\end{eqnarray}
where all functions now represent ``outer'' solutions 
valid at times beyond initial 
transients in complex formation.
We verified that this approximation holds throughout all 
of the analysis performed in the next section.

\begin{figure}[tb]
\begin{center}
\includegraphics[width=8cm]{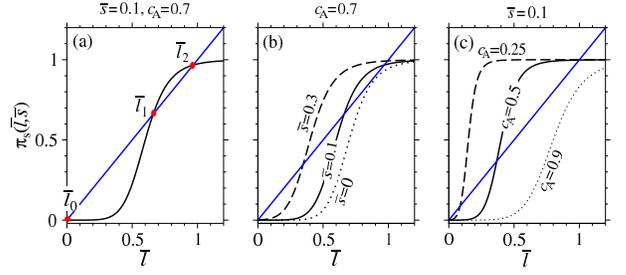}
\end{center}
\caption{Steady state configurations. (a) Graphical solution of
Eqn. (\ref{eqn:sscon}) for $\bar{s}=0.1$, and $\cA=0.7$. There are
three solutions representing two stable steady states ($\bar{l}_0$ and
$\bar{l}_2$) and one unstable equilibrium ($\bar{l}_1$).  (b) If we
fix the activation threshold ($\cA=0.7$), different values of ROS
concentration lead to different steady state configurations.
Bistability is possible if there is enough ROS, otherwise the only
stable steady state is the one with no MAPK activity.  (c) We fix ROS
concentration to $\bar{s}=0.1$ and graphically solve
Eqn. (\ref{eqn:sscon}) for different values of $\cA$.  Bi-stability
can arise only if the activation threshold is sufficiently small. The
Hill coefficient $n=8$ was used in all plots.}
\label{fig:SteadyStates}
\end{figure}

\section*{Analysis \& Results}
The spatio-temporal MAPK activation patterns can 
arise from different mechanisms. 
For example, one (or more) activation pattern could consist of 
a traveling front connecting two stable steady states, corresponding to
high and low MAPK concentrations. 
In this section we 
 describe the steady states of the system of Eqns.
(\ref{eqn:lm})-(\ref{eqn:pm}) and present an overview of the
qualitative behavior of the solutions of the model. 
After establishing the dynamics of the mathematical model, 
we use
the known model parameters to determine the nature of the MAPK patterns and 
some of their properties.

The complexity of our model requires analysis through numerical
simulations. For this purpose we use an explicit finite difference
scheme that is forward in time and centered in space, implemented in
Fortran.  We use a uniform grid discretization along the direction of
the cell layer (e.g., $x$) and a geometrical grid discretization along
the direction normal to the cell layer (e.g., $z$) to maximize
accuracy and minimize run-time. This approach and its advantages have
been previously described in \citep{Posta08}.

\begin{figure}[tb]
\begin{center}
\includegraphics[width=8cm]{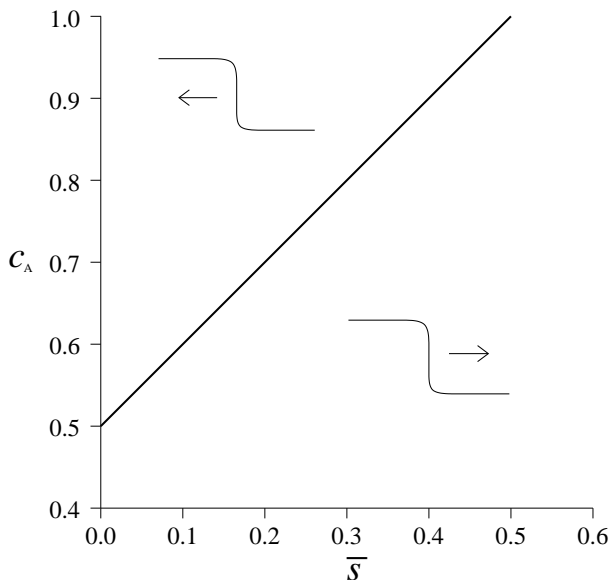}
\end{center}
\caption{Traveling waves connecting the state of MAPK activation and
the state of MAPK inactivity can move toward or away from the wound
depending on ROS concentration at cell layer level ($\bar{s}$) and
activation threshold ($\cA$) when $n \rightarrow \infty$.  The thick
black line indicates conditions under which the front is not
moving. Regimes above the line lead to traveling waves moving toward
the wound, while those below the line lead to fronts moving away from
the wound.}
\label{fig:TWdirection}
\end{figure}

\subsection*{Steady States \& Traveling Fronts}

The steady states of the model in Eqs.  (\ref{eqn:lm})-(\ref{eqn:pm})
satisfy:
\begin{equation}\label{eqn:ssgen}
\bar{l}=p, \,\, \bar{s}=\pi_{\rm s}(\bar{l},x,t), \quad\mbox{and}\quad   p=\pi_{\rm p}(\bar{l},\bar{s}),
\end{equation}
where the overbar indicates that the value of ligand or ROS
concentration is taken at $z=0$ (e.g., $\bar{l}=l(x,z=0,t)$).  The
resulting condition
\begin{equation}
\bar{l} = \pi_{\rm p}(\bar{l},\bar{s})
\label{eqn:sscon}
\end{equation}
is always satisfied by the trivial solution ($\bar{l}=0, \bar{s}=0$),
but under certain conditions it can have two more solutions as
highlighted in Fig. \ref{fig:SteadyStates}-(a). In this case the three
roots are two stable steady states, $\bar{l}_0$ and $\bar{l}_2$, and
an unstable one, $\bar{l}_1$. The two stable steady states represent a
state of no ligand signaling, $\bar{l}_0=0$, and a state of active
ligand signaling, $\bar{l}_2 > 0$, respectively. From Fig.
\ref{fig:SteadyStates} we can also infer how ROS concentration
$\bar{s}$ and the activation threshold $\cA$ control the steady states
of the system. If we fix the activation threshold at a sufficiently
high value, the system attains only the trivial steady state,
$\bar{l}=\bar{l}_0=0$, unless there is enough ROS to sustain the
signaling pathway, as shown in Fig.
\ref{fig:SteadyStates}-(b). Conversely, if we fix $\bar{s}$, the
system will be bistable only if the activation threshold $\cA$ is
sufficiently small (Fig. \ref{fig:SteadyStates}-(c)).

Bistability implies that the model admits traveling front solutions
connecting the two stable steady states.  We can approximate the front
speed analytically by considering a simpler scenario in which the
concentration of ROS at cell layer level, $\bar{s}$, is constant and
take the limit $n \rightarrow \infty$ for the Hill coefficient of
$\pi_{\rm p}$.  In this limit, the sigmoidal protease production
function $\pi_{\rm p}$ equals the Heaviside function centered at $\cA
- \bar{s}$:

\begin{equation}
\lim_{n \rightarrow \infty}
\pi_{\rm p}(\bar{l}, \bar{s})= H(\bar{l}+\bar{s}-\cA)=\left\{ \begin{array}{cc}
1 &  \bar{l} \geq \cA-\bar{s} \\  0 & \bar{l} < \cA-\bar{s} \end{array} \right. . 
\end{equation}
In this limit the system is bistable for $0 < \cA- \bar{s} < 1$, and
the roots of Eqn.  \ref{eqn:sscon} are $\bar{l}_0=0$,
$\bar{l}_1=\cA-\bar{s}$, and $\bar{l}_2=1$. Furthermore, we can
determine the velocity and direction of the traveling fronts by
proceeding as in \citep{Posta09}, obtaining:
\begin{equation}\label{eqn:vtw}
\cA-\bar{s}=\frac{1}{\pi}
\int_{\alpha v}^{\infty} \frac{\alpha \; \sqrt{q^2 - vq}}{q(1+vq)(\alpha^2 q^2 -\alpha^2 
vq+1)} \; \mbox{d}q,
\end{equation}
which gives an implicit relation for $v$, the velocity of the
traveling wave for a fixed concentration of ROS.  The integration
variable $q$ in the above integral arises from the Fourier
transformation used to derive Eqn. \ref{eqn:vtw}.  From
Eqn. \ref{eqn:vtw} we find that the front velocity is a monotonically
increasing function of the parameter $\alpha$ (see
Eqn. \ref{eqn:ndpar}).  This result is expected since an increase in
$\alpha$ corresponds to either an increase in ligand diffusivity or a
decrease in ligand binding, and both changes result in the front
reaching farther distances in a shorter amount of time.  The direction
of the front is determined by the threshold $c_A-\bar{s}$. If
$c_A-\bar{s} < 1/2$, the front of active MAPK will move away from the
wound, and deep into the cell layer. If $c_A-\bar{s} > 1/2$, MAPK
activity will recede toward the wound.  Regimes that delineate forward
and backward traveling MAPK waves are indicated in
Fig. \ref{fig:TWdirection}.  These results provide useful insight for
the general case of $n < \infty$ and diffusing ROS.  Numerical
simulations show that for Hill coefficient as low as $n=6$, the
instantaneous front velocity is within $10\%$ of that obtained from
Eqn. \ref{eqn:vtw}.  To summarize, we showed that the system can have
two stable steady states and that traveling wave solutions connecting
them are possible.  In particular, ROS can determine the existence,
velocity and direction of the fronts by effectively regulating the
activation threshold of the MAPK cascade.

\begin{table}[tb]
\begin{center}
\begin{tabular}{ccc}
%\begin{tabular}{lccc}
\hline
Parameter & 
%Description &
 Typical Value & Ref.\\
\hline
$\Dl$ & 
%\parbox[c]{2in}{\vgap Ligand diffusivity \vgap} &
 $10^{-8}-10^{-6} $ cm$^2$s$^{-1}$ & \citep{Pribyl03}\\
$\konl$ & 
%\parbox[c]{2in}{\vgap Ligand-receptor binding rate \vgap} &
 $10^{-15}-10^{-12}$ cm$^3$s$^{-1}$ & \citep{Pribyl03}\\
$\koffl$ & 
%\parbox[c]{2in}{\vgap Ligand-receptor dissociation rate \vgap} &
 $10^{-3}-10^{-2}$ s$^{-1}$ &\citep{Pribyl03} \\
$\kecl$ & 
%\parbox[c]{2in}{\vgap Ligand-receptor complex internalization rate \vgap}&
 $10^{-3}-10^{-2}$ s$^{-1}$ & \citep{Pribyl03}\\
$\kp$ & 
%\parbox[c]{2in}{\vgap Protease degradation rate\vgap} & 
$10^{-4}-10^{-3}$ s$^{-1}$ & \citep{Pribyl03}\\
$R$ & 
%\parbox[c]{2in}{\vgap Cellular surface receptor density \vgap} &
 $10^{10}-10^{13}$ cm$^{-2}$ & \citep{Pribyl03}\\
$\gp$ & 
%\parbox[c]{2in}{\vgap Protease production rate \vgap} &
 $0.17 \times 10^{8} \; $ cm$^{-2}$  s$^{-1}$ & \citep{Posta09}\\
$\gl$ & 
%\parbox[c]{2in}{\vgap Ligand production rate \vgap}& 
$0.54 \times 10^{-2}$ s$^{-1}$ & \citep{Pribyl03dis}\\
$\CA$ & 
%\parbox[c]{2in}{\vgap Protease activation threshold \vgap} &
 $10^{9}$ cm$^{-2}$ & \citep{Pribyl03dis} \\
%$v_1$ & 
%\parbox[c]{2in}{\vgap Speed of first wave \vgap}& 
%$0.27 \times 10^{-3}$ cm s$^{-1}$ & \citep{Nikolic06}\\
%$v_2$ & 
%\parbox[c]{2in}{\vgap Speed of second wave \vgap}& 
%$0.11 \times 10^{-3}$ cm s$^{-1}$ & \citep{Nikolic06}\\
%$v_3$ & 
%\parbox[c]{2in}{\vgap Speed of third wave \vgap}& 
%$0.3 \times 10^{-5}$ cm s$^{-1}$ & \citep{Nikolic06}\\
\hline
\end{tabular}
\end{center}
\caption{Typical values of model parameters.}
\label{tab:parvalues}
\end{table}

\subsection*{ROS/EGF regulation of MAPK activation}

Bistability is necessary but not sufficient for the existence of
traveling wave solutions. In this section we explore the parameter
space of the wound healing assay to determine the nature of the MAPK
fronts observed in \citep{Matsubayashi04, Nikolic06}.  To avoid
ambiguity, we divide the MAPK dynamics during wound healing into three
wave-like events. The first event corresponds to the fast activation
of MAPK initiated by the wound. It lasts until the activation front
reaches its maximum depth in the cell layer. The second event is
characterized by decrease of MAPK activity. It moves from deep into
the epithelial monolayer toward the wound. These first two events
qualitatively correspond to the first ``rebounding wave'' observed in
the experiments \citep{Matsubayashi04, Nikolic06}.  The last event
consists of a slow activation front that starts at the wound edge and
moves away from the wound.  This last ``wave'' is initiated when the
cells in the layer start moving to close the wound itself, and is
sustained when the wound is large, preventing closure in finite time
\citep{Nikolic06}.

\begin{figure}[tb]
\begin{center}
\includegraphics[width=8cm]{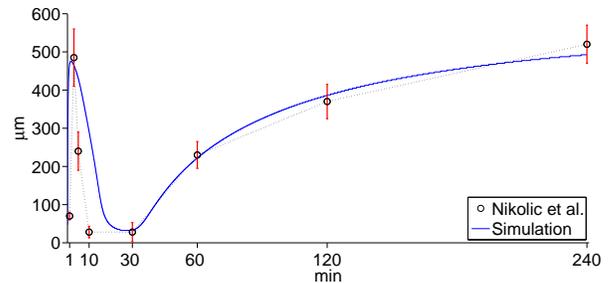}
\end{center}
\caption{Time evolution of the distance from the wound edge of the
MAPK front. The circles and the intersecting vertical bars represent
the average front position and one standard deviation error bars,
respectively. Their values have been obtained from the experimental
data in \citep{Nikolic06}.  The solid trace has been obtained through
a computer simulation of Eqns.  (\ref{eqn:lm})-(\ref{eqn:pm}) with
dimensionless parameters: $\alpha=0.35$, $\eta=16$, $\nu=5$,
$\cA=0.625$, $n=6$, $\gamma=2$, $p_A=0.975$, $m=9$, $\lambda=0.55$,
$\tau=2.56$.  These parameters are based on $\kp =5.6\times10^{-3}{\rm
s}^{-1}$, $\Dl =5.6\times10^{-7} {\rm cm}^2 {\rm s}^{-1}$, $\kecl =
\koffl =10^{-2} {\rm s}^{-1}$, $\gl =0.54 \times 10^{-2} {\rm
s}^{-1}$, $\gp =0.17 \times 10^{8} {\rm cm}^{-2}{\rm s}^{-1}$, $R=3.2
\times 10^{11} {\rm cm}^{-2}$, $\konl =10^{-15} {\rm cm}^{3}{\rm
s}^{-1}$, $\CA =10^9 {\rm cm}^{-2}$.}
\label{fig:FrontPosition}
\end{figure}

\begin{figure}[tb]
\begin{center}
\includegraphics[width=8cm]{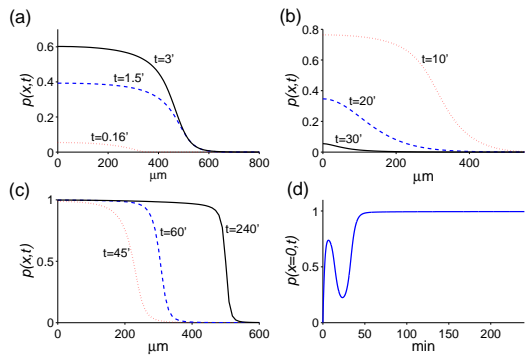}
\end{center}
\caption{Time evolution of protease. (a) After injury, the fast
diffusion of ROS drives MAPK activation. (b) As ROS diffuses away, the
positive feedback loop between MAPK and ligand is not strong enough to
sustain signaling and the depth of the front decreases. (c) Once the
cells start moving, ROS production and the ligand-protease feedback
loop fuel the activation front.  (d) Time evolution of protease
concentration at $x=0$.  The parameters used to obtain these plots are
the same as the ones used to generate Fig. \ref{fig:FrontPosition}.}
\label{fig:Waves}
\end{figure}

To reproduce the observed MAPK dynamics, we numerically integrated
Eqns.  (\ref{eqn:lm})-(\ref{eqn:pm}) using the ligand related
parameters given in Table \ref{tab:parvalues}.  Although we could not
find analogous references for physical parameters of ROS, we estimated
parameter values from various sources.  We used the self-diffusivity
of water together with the Einstein relation to bound the value of
$\eta$ between $10$ and $100$. From the results in \citep{Reynolds03}
it seems reasonable to assume $\kecs \approx \kecl$.  We assume that
the initial concentration of ligand and protease is zero, while the
concentration of ROS is represented by a narrow Gaussian with width
equal to the size of a single cell; it represents the ROS initially
released by cell rupture.  Our numerical results are summarized in
Figs. \ref{fig:FrontPosition} and \ref{fig:Waves}.
Fig. \ref{fig:FrontPosition} compares the time evolution of the
distance of the front from the wound as predicted by
Eqns.(\ref{eqn:lm})-(\ref{eqn:pm}) with the experimental values
observed in \citep{Nikolic06} for scratch wounds.  The position of the
front is determined by evaluating the inflection point of protease
concentration after each time step.  The model is able to replicate
the observed MAPK behavior and we used it to investigate the dynamics
of the three activation wave-like patterns. Fig. \ref{fig:Waves} shows
the time evolution of the profiles of the three MAPK ``waves''.  The
first wave (Fig. \ref{fig:Waves}-(a)) is driven by ROS production at
the onset of wound and its fast diffusion.  However, there is not
enough ROS for either ligand or protease concentration to reach the
signaling steady state (e.g., $p=\bar{l}=1$).  As a result, the first
``wave" can only propagate as far as $\sim 480\mu m$ before receding.
As ROS diffuses away, protease concentrations decrease
(Fig. \ref{fig:Waves}-(b)) until the cells start to move (after about
30 minutes from injury). At that time, ROS is produced by the moving
cells and fuels the positive feedback loop between ligand and
protease. The nonlinear effects of $\pi_{\rm p}$ allow protease levels
to increase until they reach a signaling steady-state. At this point,
the front moves like a true traveling-wave (Fig. \ref{fig:Waves}-(c)),
with its speed and distance traveled regulated by the ROS source
function $\pi_{\rm s}$. We can also track the time-evolution of the
variables in the model. Fig. \ref{fig:Waves}-(d) shows how protease
concentration at the wound edge $x=0$ changes in time.  From this
graph we observe that during the first MAPK event the protease
concentration never reaches the ``signaling'' steady-state $p \approx
1$ and eventually decreases. During the third wave, protease
concentration reaches the ``signaling'' steady-state and remains there
as shown by the flat part of the graph in Fig. \ref{fig:Waves}-(d).
To summarize, only the third MAPK front behaves as a true traveling
wave, while the initial two events (corresponding to the first
``rebounding wave'' observed in experiments) are actually transient,
diffusion-driven patterns.

\section*{Conclusions}

We have formulated a mathematical model for the dynamics of
intercellular signaling observed during wound healing
experiments. From this model, we were able to replicate the signaling
patterns observed in \citep{Matsubayashi04,Nikolic06}, and to provide
insight regarding their nature.  Our choice of EGF as signaling ligand
is based upon literature review, but lacks experimental evidence
within the epithelial wound healing assay. However, we showed that the
properties of EGF (and EGFR) fit the profile for the unidentified
diffusible signals mentioned in \citep{Nikolic06}. Our model can be
expanded to incorporate other diffusible signaling molecules. Although
it may be possible to find physiologically realistic sets of
parameters that lead to signaling patterns consisting of three
separate traveling waves, the parameters associated with the
EGF/ROS/MAPK system lead to only one final traveling wave. The first
two notable events being described by purely diffusive and decays
dynamics, respectively.
 
An aspect of our current model that needs improvement and further
analysis is the determination of the ROS source function $\Pi_{\rm
S}$.  From the experiments in \citep{Nikolic06}, this term seems to be
negligible since they only detected the presence of extracellular ROS
up to $\sim 10$min after wounding.  However, there is substantial
evidence indicating that cell motility and ligand-receptor binding can
induce ROS production.  A plausible explanation for these conflicting
results could be that ROS produced after wounding is fully recaptured
by intracellular processes (including EGFR phosphorylation) and never
crosses the cell membrane.  Nonetheless, we performed many numerical
tests and found that if $\Pi_{\rm S}=0$ (data not shown), then all
three MAPK events are diffusion driven and the signaling pattern is
due to the different diffusion properties of ROS and ligand (EGF).  A
more physically realistic approach might be to include the mechanical
events that take place within the cell layer and the reaction that
lead to ROS production after ligand binding.  Such an approach could
provide important insights about the mechanisms of post -wound ROS
production and their relevance within the MAPK signaling context.\\

\section*{Acknowledgments}
We thank M. Gibbons, S. Shvartsman, and C. Muratov for useful
discussion.  This work was supported by NSF grant DMS-0349195 and NIH
grant K25 AI58672.

\bibliography{PostaChouJTB}

\end{document}